# Multimode Fiber Imaging Based on Hydrogel Fiber

Lele He,[1†] Mengchao Cao,[1†] Lili Gui,[1] Jingjing Guo,[2] Xiaosheng Xiao[1]*

[1] State Key Laboratory of Information Photonics and Optical Communications, School of Electronic Engineering, Beijing University of Posts and Telecommunications, Beijing 100876, China

[2] School of Instrumentation and Optoelectronic Engineering, Beihang University, Beijing 100191, China

**ABSTRACT** We demonstrate a multimode fiber imaging technique based on hydrogel fibers, which are suitable for biomedical applications owing to their biocompatibility and environmental friendliness. High-resolution handwritten images are successfully recovered by utilizing a Pix2Pix image generation network.

## I. Introduction

Multimode fiber (MMF) imaging technology holds broad application prospects in bio-imaging, industrial detection, and other fields, and has attracted extensive attention in the biomedicine and industry. Due to modal dispersion and mode coupling within MMFs, speckle patterns are generated when images are transmitted through MMFs. As an effective tool, neural networks can restore original images from speckle pattern via their powerful nonlinear modeling capabilities, end-to-end learning mechanisms, and strong adaptability and generalization abilities [1].

In the biomedical field, the transmission of light signals in biological tissues is limited by the intrinsic light scattering and absorption. The traditional silica fibers are highly stiff and rigid, which may easily cause damage to host tissues after implantation. This issue can be addressed by developing specialty optical fiber with biocompatibility and environmental friendliness. Specialty optical fibers have been proposed for biochemical sensing. Guo et al. used polyethylene glycol diacrylate (PEGDA) hydrogel and polydimethylsiloxane to fabricate waveguide-based sensors with diverse optical and mechanical properties [2]. Wu et al. explored methods for the preparation of soft and malleable optical waveguides using temperature - and pH-sensitive polymer materials, opening up new possibilities for biomedical applications [3]. Specialty fibers have also been introduced into imaging by Shan et al. [4]. They developed a flexible, biodegradable step-index fiber based on citrate for fiber imaging. Three letters (P, S, and U) are projected on the near end of the fiber and random speckles are recorded on the far output. The input image was recovered from the speckle by using the pre-measured impulse response combined with the least square method, demonstrating the potential of citric citrate based polymer fiber for image transmission. However, the resolution (8×8) of the imaging is not enough to be used in practical applications.

In this Presentation, we will report the exploration of the hydrogel-based MMF imaging technology. The non-toxic and biocompatible hydrogel fiber we made is used as the imaging-transmission medium. A hydrogel optical fiber imaging system was built for capturing speckles. We optimized a Pix2Pix network model to achieve the reconstruction of the speckle patterns output from the hydrogel MMF, which contains a large number of modes. Handwritten dataset MNIST with high-resolution was successfully reconstructed by this neural network model. This work lay a foundation for the application of specialty waveguides in biomedical imaging.

## II. Principle and method

To fabricate PEGDA hydrogel fibers, as shown in Fig. 1, 40% w/v PEGDA and 2% w/v photopolymerization initiator (2-hydroxy-2-methylphenylacetone) were thoroughly mixed in deionized water using a vortex mixer to prepare a precursor solution. The precursor solution was injected into the waveguide mold using a disposable syringe for an appropriate time at rest, and then irradiated with a UV curing lamp (wavelength 365 nm, power density 5 mW/cm$^2$) for 5 minutes. The precursor solution underwent photocrosslinked under the action of photopolymerization initiator, and the formed hydrogel fiber was obtained after demolding. The detailed characteristics and parameters of this type of hydrogel fiber can be found in Reference [5].

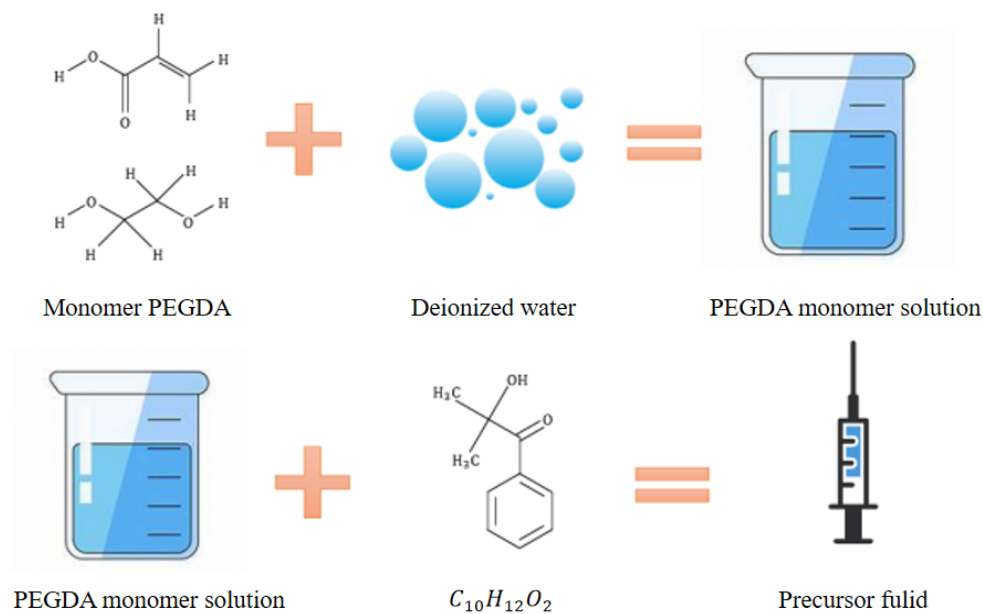

(a)

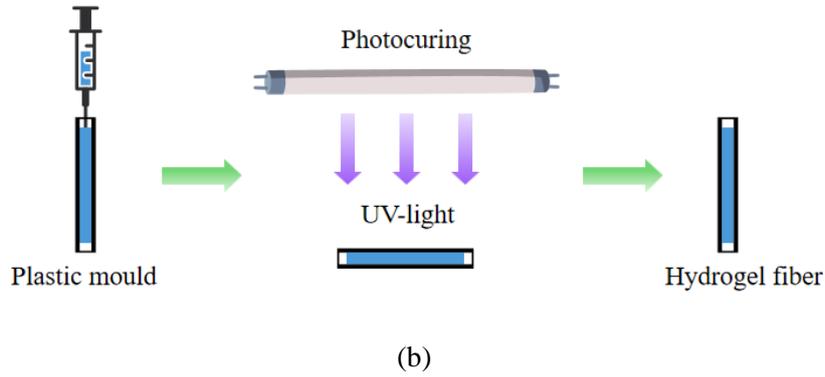

(b)

Fig. 1. (a) Reagents used to prepare the hydrogel precursor, (b) Schematic of the hydrogel fiber preparation process.

We then analyzed and optimized the optical coupling efficiency and transmission loss of the hydrogel fiber to ensure its effective application in biomedical imaging. Coupling efficiency, a key index in evaluating the performance of optical fiber systems, directly affects signal quality in imaging systems. To enhance the efficiency of optical coupling, a silica MMF jumper was connected to one end of the hydrogel before solidify the hydrogel fiber, as shown in Fig. 2. During the applications of imaging, this silica MMF is placed at the distal end, and the illumination light is coupled into the MMF jumper. This protects the hydrogel fiber from damage or water loss, thus facilitating experimental operation and practical use. Additionally, to prevent the hydrogel fiber from water loss in our experiments, silicone tubes were employed to encase the fiber, and deionized water was injected to slow the water evaporation. Furthermore, by adjusting preparation conditions such as PEGDA concentration, UV light curing time, and light intensity, we reduce the transmission loss and enhance the mechanical properties of the hydrogel fiber, thereby optimizing its optical characteristics and transmission efficiency.

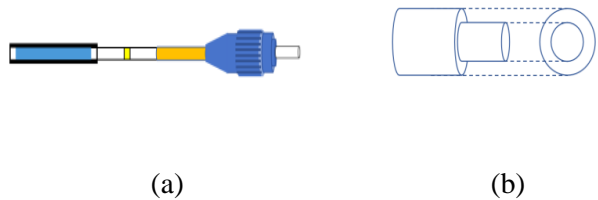

(a)              (b)

Fig. 2. (a) Schematic of the hydrogel fiber connected to a silica MMF jumper.   (b) Hydrogel fiber wrapped in a silicone tube.

Then we built an MMF imaging system, which was similar to that of our previous work [1]. Herein the hydrogel fiber with the fused jumper served as the MMF, as shown in Fig. 3(a). Compared with traditional silica MMFs (e.g., as in Ref. [1]), the speckle patterns acquired via the hydrogel fibers are smaller and denser, as shown in Fig. 3(b). This phenomenon arises from the larger core diameter of the hydrogel fiber than silica MMFs, enabling it to accommodate much more modes, which introduces larger modal dispersion and more complex coupling. This unique feature makes it much more difficult to recover the images than the case with traditional silica MMF.

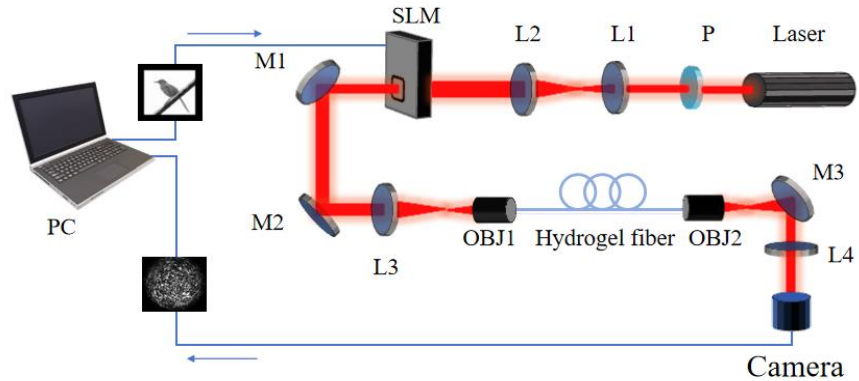

(a)

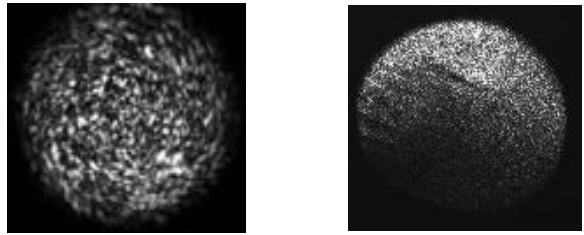

(b)

Fig. 3. (a) Hydrogel fiber imaging system. P: half waveplate; L1-L4: Lens; SLM: Spatial light modulator; M1-M3: mirror; OBJ1,2：micro objective; MMF: multimode fiber) (b) Speckle from traditional silica MMF (left), Speckle from hydrogel fiber (right).

We attempted image reconstruction tasks using classical convolutional neural networks, such as CVNN, VGG and Resnet [1], but the results were unsatisfactory, as shown in Fig. 4. After experimenting with various networks, we found that Pix2Pix yielded superior performance. The Pix2Pix network is specifically designed for image-to-image conversion tasks. Typically, datasets for such models are divided into two groups, each representing different feature styles. By learning the mapping relationship between these groups of images, the network can be utilized for tasks such as image style conversion, image coloring, and image restoration. To enhance the Pix2Pix network's applicability in processing the speckle patterns from hydrogel fiber and to improve its efficiency in handling complex information, we proposes several improvements to the network model. These enhancements include increasing the number of convolutional kernels and feature map channels, utilizing grayscale maps as input data, and adjusting the network structure. These modifications render the network better suited for processing complex information, thereby enhancing its processing capacity and operational efficiency. This holds particular significance for recovering the speckle patterns from hydrogel fiber.

III. Experimental Results

In the hydrogel fiber imaging experiment, the MNIST handwritten dataset was employed as input images, resulting in a collection of 14,000 sets of raw image-speckle datasets, of which 80% were allocated for training and 20% for testing. Prior to network training, preprocessing of the speckle images was conducted, involving cropping and image enhancement, aimed at achieving higher contrast and clearer data. For details on image preprocessing, models and parameters, etc., please refer to the supplementary material. Fig. 4 shows the typical reconstruction results with different neural networks.

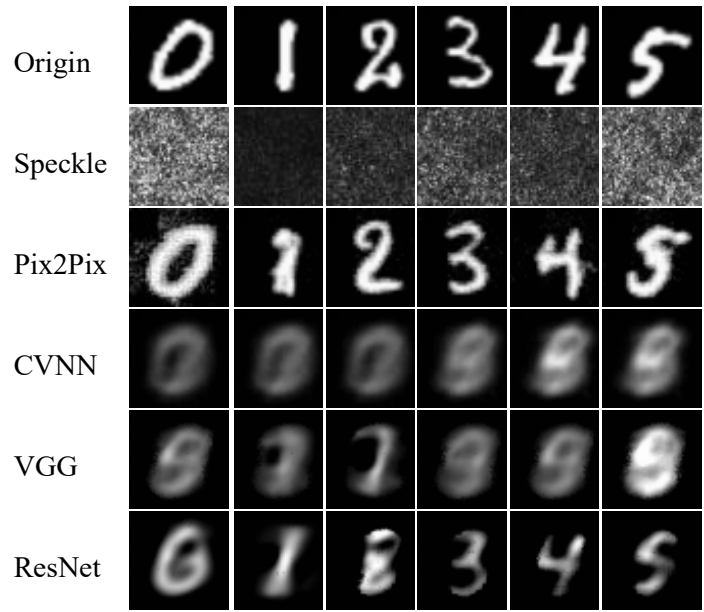

Fig. 4. Reconstruction results of hydrogel fiber based imaging system, using four different neural network models.

It can be seen from Fig. 4 that, the performance of CVNN network model and VGG network model is very poor, and basically cannot reconstruct the images. Due to the principle of ResNet, the network depth of ResNet model is much higher than the above two models, so it has a stronger ability to extract high-dimensional abstract features in speckle. Although the reconstruction results of ResNet are fuzzy, the numbers in the recovered image can be roughly recognized. Unlike these three traditional models described above, the Pix2Pix network model can basically complete the speckle reconstruction of the handwritten digital data set, demonstrating better performance in recovering the images of hydrogel fiber.

The reason why Pix2Pix network performs better in our image restoration is as follows: The Pix2Pix network is specifically designed for image-to-image translation tasks, thus having inherent advantages in handling image reconstruction and conversion. In contrast, CVNN, VGG, and ResNet are generally more suitable for image classification and recognition tasks; when applied to image reconstruction, they require more manual adjustments and optimizations. In the Pix2Pix network, the generator and discriminator compete with each other through adversarial training. This training mechanism helps the generator produce more realistic and clear images when dealing with speckles and noise in multimode fiber imaging, thereby improving the quality of image reconstruction. The Pix2Pix network enables end-to-end training, directly learning the mapping relationship from input images to output images, which allows it to better capture the complex relationships between inputs and outputs. In addition, regarding the selection of loss functions, traditional networks such as VGG and ResNet usually adopt mean squared error loss, which tends to focus on global error minimization and may lead to deficiencies in local details of imaging results. However, the Pix2Pix network combines adversarial loss and L1 loss during training, guiding the generator to produce images closer to the real data distribution while striving to reduce pixel-level errors. This achieves favorable optimization both in overall performance and details, and helps enhance the model's robustness to noise and outliers.

We further attempted to reconstruct the natural dataset ImageNet using hydrogel fibers, but the reconstruction results were unsatisfactory so far, as shown in Fig. 5. This is mainly because the natural image contains a much richer levels of grayscale and more complex details, making it more challenging to recover compared to the handwritten dataset.

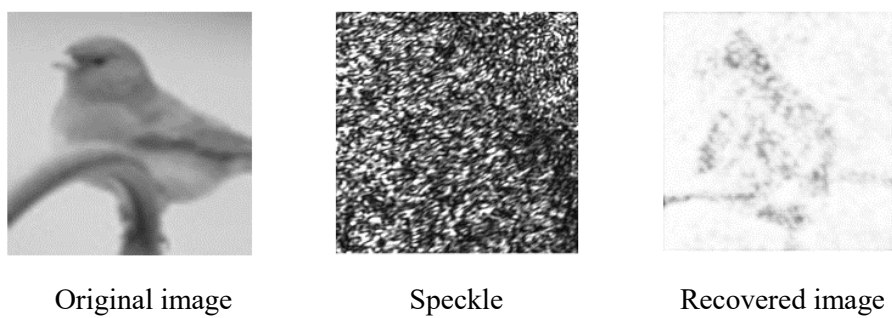

      Original image            Speckle            Recovered image

Fig. 5. Reconstruction results of the ImageNet dataset using the current Pix2Pix model.

Overall, the Pix2Pix model has shown the potential to reconstruct speckle images into original images. In the future, the quality of reconstructed images can be further improved by expanding training data, optimizing model architecture, introducing post-processing steps and adopting more complex training strategies.

## IV. Conclusion

This Presentation reported the MMF imaging based on hydrogel fibers. The preparation, fabrication, and physical properties of PEGDA hydrogel MMF are presented. Hydrogel MMF imaging is demonstrated, by optimizing the structure and parameters of neural network model. The speckle image reconstruction of handwritten data set MINSIT through hydrogel fiber is realized successfully. Furthermore, the reconstruction of natural images from hydrogel fiber is discussed. This work lays a research foundation for the application of special hydrogel fiber in the field of biomedical imaging.

**Acknowledgements**

This work was supported by the National Natural Science Foundation of China (No. 62375024), Fundamental Research Funds for the Central Universities from China (No. ZDYY202102-1), Fund of State Key Laboratory of Information Photonics and Optical Communications (Beijing University of Posts and Telecommunications) of China (Nos. IPOC2021ZR02 and IPOC2020ZT02).

†These authors contributed equally to this work.

*Corresponding author. Email: xsxiao@bupt.edu.cn

# Supplementary material

## 1. Optical fiber

To reduce transmission loss, the preparation conditions of the hydrogel fiber, such as photocuring time, were optimized in the experiment. The curing time adopted for fiber fabrication was 70 seconds. The manufactured hydrogel fiber can be cut with a thin blade to yield a relatively intact end face, and its normal service life in air has been extended from 1 hour to approximately 7 hours.

Figure S1 presents a photograph showing the connection between a conventional silica multimode fiber and the hydrogel fiber.

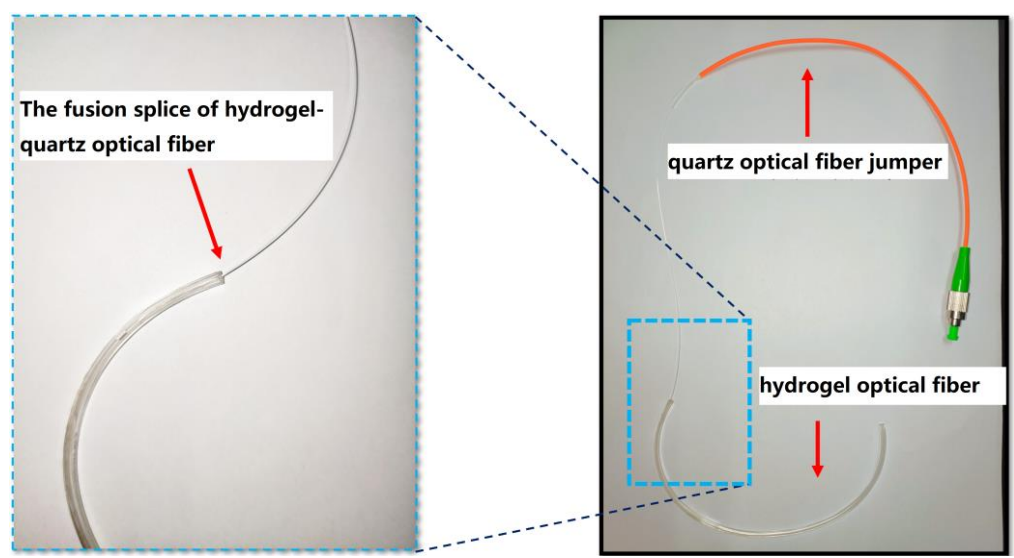

Figure S1 Connection between a conventional silica multimode fiber and the hydrogel fiber.

The right side of Figure 3(b) shows a typical speckle pattern collected. It can be seen from the figure that the obtained speckle pattern is not a perfect circle, which is because the pressing block of the V-shaped holder will squeeze the optical fiber to a certain extent during fixing. It can also be observed in the figure that there are irregular cracks on the end face of the hydrogel optical fiber, and the reasons for their formation are as follows: Unlike traditional silica fibers, hydrogel optical fibers have high flexibility and elasticity. When an optical fiber cleaver is used to cut the hydrogel optical fiber, the hydrogel optical fiber will first undergo significant deformation under the pressure of the blade instead of breaking immediately, until it is squeezed and crushed. In this experiment, a thin blade was used to manually cut the hydrogel optical fiber. During the cutting process, the hydrogel optical fiber was first placed on a clean and flat operating table to avoid bending near the cutting position. Then, the blade was held to cut vertically and at a constant speed at the appropriate position, so as to obtain a possibly complete optical fiber end face.

## 2. Image Preprocessing

After collecting speckle images, preprocessing is required before neural network training, including image cropping, image enhancement, and data normalization.

(1) Image Cropping: Cracks in speckle patterns can interfere with feature extraction during neural network training, reducing the distinguishability of reconstruction results. Thus, crack-free central regions are cropped as the dataset (using 256×256). Since the central area contains most of the original light field information coupled into the fiber (as concluded in our previous investigations), the cropped speckles suffice for image reconstruction.

(2) Image Enhancement: Hydrogel fibers have higher loss than traditional silica fibers, resulting in dim speckle patterns due to low light signal intensity. According to our previous study on noise impact, linearly increasing speckle brightness within a certain range can improve contrast, sharpness, and feature clarity, thereby enhancing reconstruction quality. Thus, brightness of cropped speckles is linearly increased for subsequent training.

(3) Data Normalization: It balances the contributions of all features to the loss function, accelerating model convergence. By limiting data to a small range, it enhances training stability and avoids gradient explosion/vanishing. It also helps the model learn generalized features, improving generalization ability for new datasets and facilitating practical deployment. Additionally, it reduces feature correlation in multimode fiber imaging, enhancing model interpretability by enabling clearer learning of independent feature information.

## 3. Pix2Pix Model

After preprocessing the collected images, our attempts showed that traditional neural network models yield unsatisfactory results in reconstructing speckles from hydrogel optical fibers due to their insufficient feature extraction capability. Thus, the Pix2Pix model was adopted. This model focuses on image generation and can produce results more similar to the original images based on information in speckles by controlling the adversarial loss function and L1 loss function.

The Pix2Pix network is specifically designed for image-to-image translation tasks. Datasets for such models are typically divided into two groups with distinct feature styles. By learning the mapping relationship between the two groups of images, it can be applied to scenarios like image style transfer, image coloring, and image inpainting. For example, in the task of redrawing real photos into cartoon-style images, one group may be real portraits, and the other their corresponding cartoon versions. In this study, the core task is to realize the inverse reconstruction of speckles in the test set by learning the mapping relationship between speckles and original patterns in the training set. Essentially, the Pix2Pix network generates new images, similar to generative networks, and also consists of two core components: a generator and a discriminator. The generator generates output images from input images, while the discriminator distinguishes between generated images and real ones. During training, the generator and discriminator compete adversarially to improve the authenticity of generated images. To enhance training efficiency, the Pix2Pix network usually employs a U-Net architecture as the generator.

The structure of the generator model is shown in Figure S2, which labels the name of each layer and the dimensions of input and output data. The first parameter of the data dimension is the batch size, set here as "?" to indicate that the model can dynamically adjust to adapt to inputs of different batch sizes. The next two parameters represent the dimensions of input data, i.e., the width and height of the input image, and the last parameter denotes the number of channels. In this generator model, the input grayscale image is 256×256 with 1 channel. Due to the large number of modes in hydrogel optical fibers, more encoding layers are needed to extract sufficient features. This experiment used 7 downsampling layers; each layer doubles the number of feature map channels while halving the spatial size of the feature map. In the decoding stage, 6 upsampling layers are used to gradually restore the spatial size of the image. Meanwhile, skip connections link the decoded feature maps with the corresponding-sized feature maps obtained from previous downsampling. Finally, a transposed convolution layer outputs an image with the same resolution as the input. In the data preprocessing step, the grayscale matrices of the speckle dataset were normalized to constrain input values within the range [-1, 1], so the tanh activation function with a value range of (-1, 1) was selected.

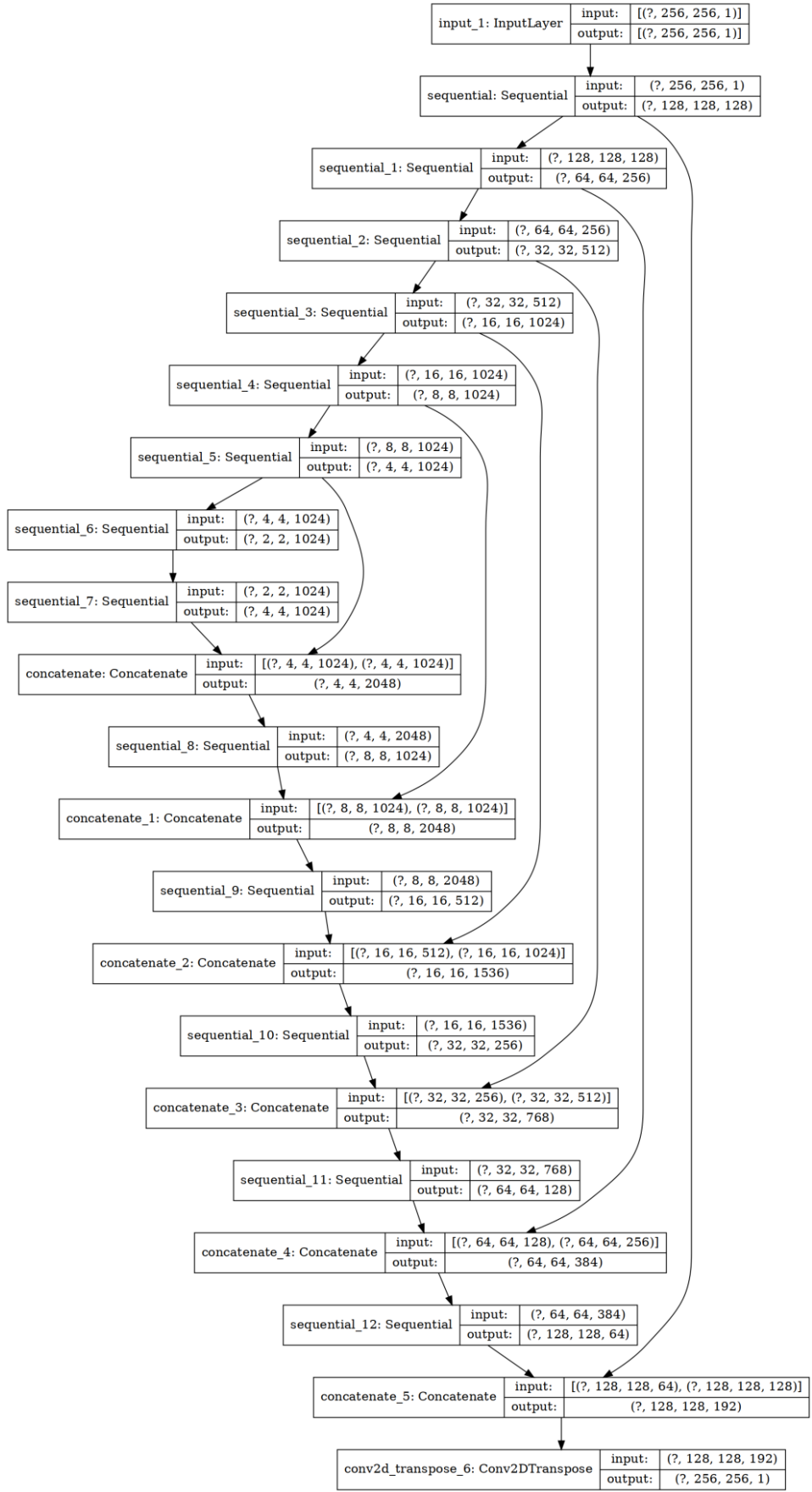

Figure S2 Structure of the generator model

The structure of the discriminator model is shown in Figure S3. Its inputs are real images from the dataset and generated images from the generator. A series of downsampling layers reduce the resolution while increasing feature depth, and the final output is a probability evaluating whether the input image is real or generated by the generator. Each (1, 1) region in the final output feature map corresponds to a (256/30, 256/30) region in the input image, approximately (8.53, 8.53) pixels. Due to the overlapping receptive fields of convolutional layers, each output element represents aggregated information from a larger region in the input image. This structure enables the discriminator to evaluate the authenticity of the combination of input and target images within a larger receptive field. In this fully convolutional discriminator model, downsampling layers with a stride of 2 reduce each dimension of the feature map while increasing the receptive field (the area of the input image "seen" by the network). The last two ZeroPadding2D layers and convolutional layers with a stride of 1 fine-tune the size of the feature map, ultimately adjusting the output feature map to 30×30. This design allows the discriminator network to evaluate not only the overall authenticity of the image but also the local authenticity of various small regions.

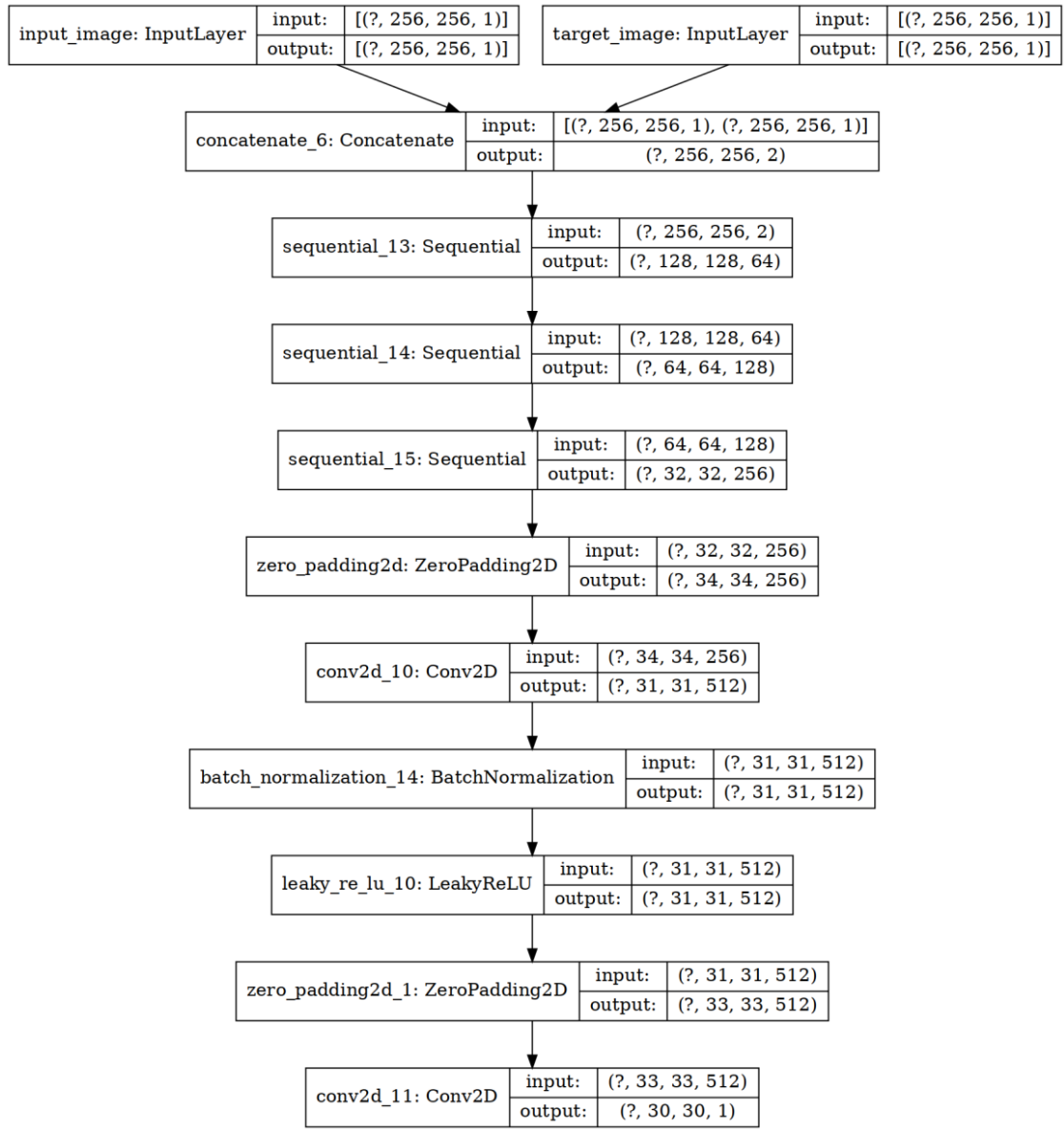

Figure S3 Structure of the discriminator model

Feature maps are obtained by convolving input images with convolution kernels and transmit information between layers of the neural network. They can be regarded as multi-dimensional arrays; the number of convolution kernels determines their number of channels, with each channel specifically encoding a certain feature or pattern of the input data. As the number of convolution kernels increases, the number of feature map channels also increases. In deep network structures, each channel of the feature map represents more abstract high-level features such as edges, colors, and textures. The number of feature map channels directly affects the number of model parameters: more channels mean the network can capture richer features but also increase computational resource consumption. Typically, a color image has channels corresponding to its color channels (i.e., RGB three channels), while a grayscale image has only one corresponding grayscale channel. In this experiment, hydrogel optical fibers have more modes and can carry more information than traditional quartz optical fibers. To enable the network to handle the task of reconstructing original images from speckles containing complex information, we adjusted the number of convolution kernels in the generator and discriminator to increase the number of feature map channels, thereby extracting more features from speckle images. Meanwhile, grayscale images were used as input to reduce the number of network parameters and accelerate computation.

**4. Speckle Reconstruction Process of Hydrogel Multimode Fiber**

In the hydrogel multimode fiber imaging experiment, a total of 14,000 groups of original image-speckle datasets were collected, with 80% used as the training set and 20% as the test set. Before training the Pix2Pix network, data preprocessing was performed using the method described in the previous section, followed by loading the dataset for neural network training. The parameter settings of the Pix2Pix network in this experiment are elaborated below.

For the downsampling part of the generator: each convolution kernel is set to 4×4 with a stride of 2, and "same" padding is used to keep the output dimension consistent with the input. In the neural network, as depth increases, the distribution of input data in each layer changes. After convolution, Batch Normalization is applied to standardize the mean and variance of the data, ensuring a stable distribution before input to the next layer, which alleviates gradient vanishing/explosion and accelerates training convergence. Finally, the LeakReLU function is used for nonlinear activation.

For the upsampling part of the generator: the convolution kernel is also 4×4 with a stride of 2, and "same" padding is adopted for transposed convolution. After Batch Normalization, Dropout is applied—randomly deactivating 50% of neurons (setting their weights to 0) before input to the next layer. This ensures that only 50% of activated neurons propagate and update weights during each training iteration, reducing the model's over-reliance on specific samples, promoting learning of general features, and improving generalization ability. Additionally, it reduces network parameters and speeds up training.

The discriminator's input layer receives both generated images from the generator and real images, concatenating them to consider their features simultaneously. Three downsampling convolution layers with 64, 128, and 256 kernels respectively (stride=2) are used for feature extraction. Zero-padding convolution is then applied to maintain feature map size, followed by Batch Normalization and LeakReLU for nonlinear activation. Finally, a 1-channel convolution layer is used as the output layer to represent the probability that the input image is real.

The generator's loss function consists of GAN loss and L1 loss. The GAN loss is expressed as:

$$L_{GAN}(G) = -E_x[logD(G(x))] \qquad (S1)$$

where *x* is the input speckle image, *G(x)* is the image generated by the generator, and *D(G(x))* is the probability that the discriminator identifies the generated image as real. The adversarial loss represents the mathematical expectation of the logarithmic probability that the discriminator judges the generated image as real. The L1 loss constrains the pixel-level similarity between the generated and original images, expressed as:

$$L_{L1}(G) = E_{x,y}\left[\left\|y - G(x)\right\|\right] \tag{S2}$$

where *y* is the original image, and *G(x)* is the speckle-generated image.

The total generator loss function is:

$$L_G = L_{GAN}(G) + \lambda L_{L1}(G) \tag{S3}$$

where $\lambda$ is a weighting coefficient, set to 200 in this experiment.

The discriminator's loss function also includes two parts:

$$L_D = -E_{x,y}[logD(x, y)] - E_x[log(1 - D(x, G(x)))] \tag{S4}$$

where *x* is the input speckle image, *y* is the corresponding original image, and *G(x)* is the generated image. The first part represents the discriminator's judgment on the "speckle-original image" pair, and the second part on the "speckle-generated image" pair.

During training, the generator and discriminator parameters are updated alternately: first, fix the generator and update the discriminator to minimize $L_D$ (enhancing its ability to distinguish real and generated images), then fix the discriminator and update the generator to reduce $L_G$. The Adam optimizer is used for both, with a learning rate of 0.001 and $\beta_1 = 0.5$. This process is iterated until network convergence, enabling the generator to produce images similar to the original, thus achieving speckle reconstruction.